\documentclass{aa}  

\usepackage{graphicx}	
\usepackage{amsmath}	
\usepackage{amssymb}	
\usepackage{xcolor}     
\usepackage{soul}       

\newcommand{\Msun}{M_{\odot}}
\newcommand{\MBH}{M_{\rm bh}}
\newcommand{\JAVELIN}{{\tt JAVELIN}}
\newcommand{\JAVELINEXT}{{\tt JAVELIN-Ext}}
\newcommand{\PyCS}{{\tt PyCS}}
\newcommand{\CREAM}{{\tt CREAM}}
\newcommand{\ugrizy}{{\it ugrizy}}
\newcommand{\LSSTu}{{\rm LSST-{\it u}}}
\newcommand{\LSSTgrizy}{{\rm LSST-{\it grizy}}}
\newcommand{\tlag}{t_{\rm lag}} 
\newcommand{\DTmean}{\left< \delta t \right>_\lambda}
\newcommand{\DTmeanLSSTu}{\left< \delta t \right>_\LSSTu}
\newcommand{\bd}{\begin{displaymath}}
\newcommand{\ed}{\end{displaymath}}
\newcommand{\be}{\begin{equation}}
\newcommand{\ee}{\end{equation}}
\newcommand{\beaa}{\begin{eqnarray*}}
\newcommand{\eeaa}{\end{eqnarray*}}
\newcommand{\bea}{\begin{eqnarray}}
\newcommand{\eea}{\end{eqnarray}}

\newcommand{\sref}[1]{Section~\ref{#1}}
\newcommand{\aref}[1]{Appendix~\ref{#1}}
\newcommand{\fref}[1]{Figure~\ref{#1}}
\newcommand{\fsref}[1]{Figures~\ref{#1}}
\newcommand{\tref}[1]{Table~\ref{#1}}
\newcommand{\eref}[1]{Equation~(\ref{#1})}

\begin{document} 

\title{Twisted quasar light curves: implications for continuum reverberation mapping of accretion disks}
\titlerunning{Twisted quasar light curves}

\author{
J.~H-H.~Chan\inst{\ref{epfl}}\and
M.~Millon\inst{\ref{epfl}}\and
V.~Bonvin\inst{\ref{epfl}} \and
F.~Courbin\inst{\ref{epfl}}  
}

\institute{
Institute of Physics, Laboratory of Astrophysics, \'Ecole Polytechnique F\'ed\'erale de Lausanne (EPFL), Observatoire de Sauverny, 1290 Versoix, Switzerland 
\label{epfl}
\goodbreak
}

\date{\today}

\abstract{
With the advent of high-cadence and multi-band photometric monitoring facilities, continuum reverberation mapping is becoming of increasing importance to measure the physical size of quasar accretion disks. 
The method is based on the measurement of the time it takes for a signal to propagate from the center to the outer parts of the central engine, assuming the continuum light curve at a given wavelength has a time shift of the order of a few days with respect to light curves obtained at shorter wavelengths.
We show that with high-quality light curves, this assumption is not valid anymore and that light curves at different wavelengths are not only shifted in time but also distorted:
in the context of the lamp-post model and thin-disk geometry, the multi-band light curves are in fact convolved by a transfer function whose size increase with wavelength. 
We illustrate the effect with simulated light curves in the LSST \ugrizy\ bands and examine the impact on the delay measurements when using three different methods, namely \JAVELIN, \CREAM, and \PyCS.
We find that current accretion disk sizes estimated from \JAVELIN\ and \PyCS\ are underestimated by $\sim30\%$ and that unbiased measurement are only obtained with methods that properly take the skewed transfer functions into account, as the \CREAM\ code does.
With the LSST-like light curves, we expect to achieve measurement errors below $5\%$ with typical 2-day photometric cadence. 
}


\maketitle


\section{Introduction}
Active Galactic Nuclei (AGNs) are astrophysical sources powered by the accretion of hot gas onto super-massive black holes (SMBHs) at the center of galaxies. 
Gas or dust around a SMBH orbits in a plane around the SMBH center, forming a so-called accretion disk. 
In current models, the central accretion disk is considered to be optically thick and geometrically thin \citep{Shakura&Sunyaev73}. 
Its emission is a combination of the internal heat from viscous dissipation and external heat from reprocessing of the UV/X-ray source near the SMBH. 
Considering that the disk luminosity is produced by black-body radiation, the temperature profile at large distance, $R$, follows $T\propto R^{-3/4}$.

Understanding the growth and evolution of SMBH in AGNs requires to study the structure of their accretion disk. 
Currently, size measurements are carried out either using microlensing in lensed AGNs \citep[e.g.][]{Schechter&Wambsganss02,Kochanek04,MorganEtal10,MorganEtal18}, or reverberation mapping \citep[e.g.][]{FausnaughEtal16,JiangEtal17,MuddEtal18,YuEtal18}. 
The basic idea of reverberation mapping is to measure the time lag $\tau_{\rm lag}$ between broad-line and continuum fluxes from spectroscopic monitoring \citep{Blandford&Mckee82}.
Assuming that the broad line emission is triggered by the central emission, the lag is considered as the light-travel time from the central illuminating source to the Broad-Line Region (BLR), i.e. $R_{\rm BLR}=c \tau_{\rm lag}$ and provides a measurement of BLR's size. 
Similarly, it is natural to use the same method to measure the disk size itself, since the continuum emission across the disk is also driven by the central source.

\begin{figure*}[t!]
\centering
\includegraphics[scale=0.6]{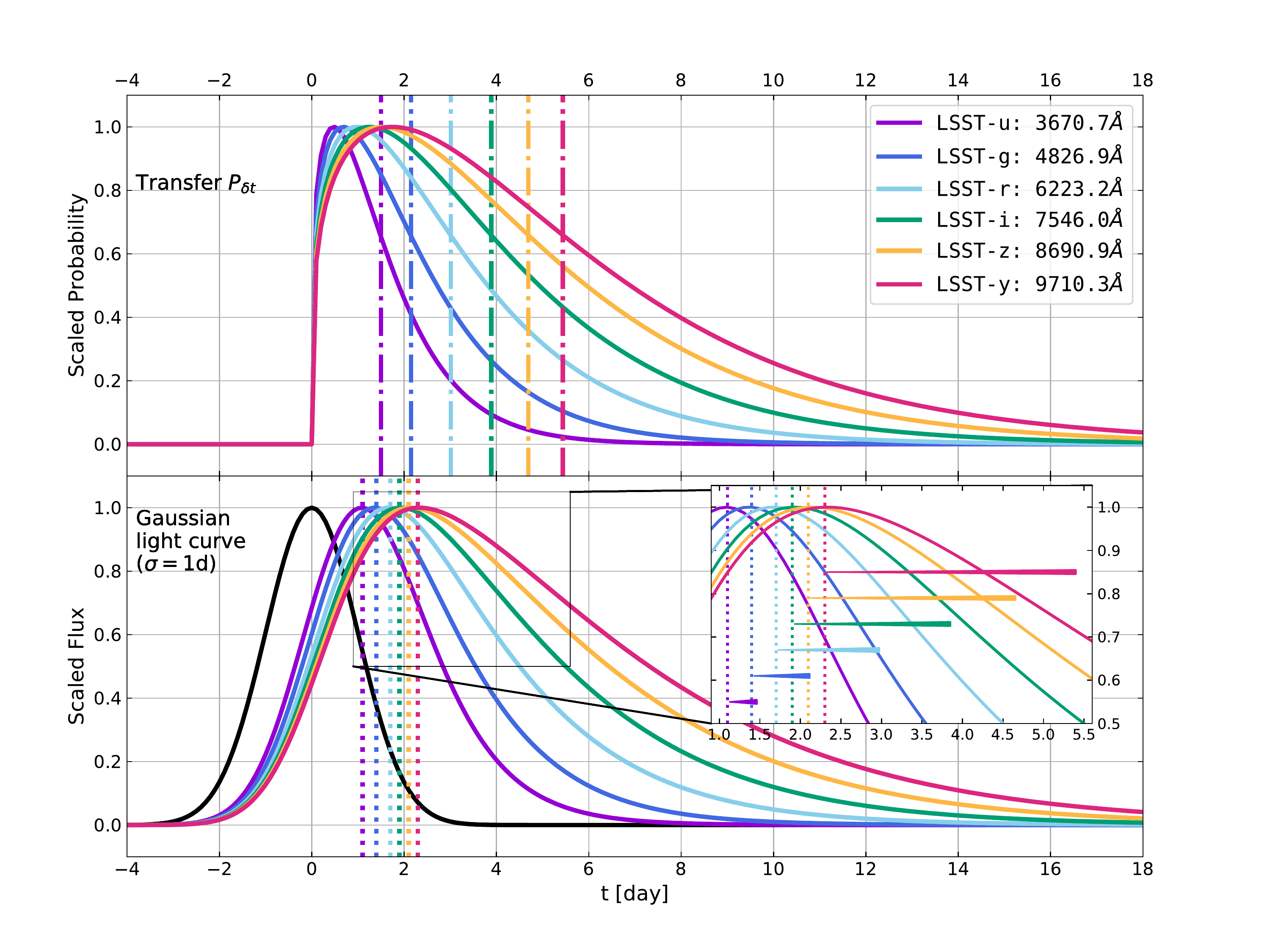}
\caption{{\it Top:} Transfer functions for the LSST filters. The mean delays $\DTmean$ of each transfer function \eref{eqn:mean_delay} are represented as dot-dashed vertical lines. {\it Bottom:} Light curves generated using a single Gaussian as a driving source (shown in black). The peak positions are represented as vertical dotted lines. 
The inset highlights with colored wedges the difference between the peak position and the mean delay for each band.
}
\label{fig:SimGau}
\end{figure*}

Early studies have obtained continuum lags for several targets, in particularly using the {\it Swift} data \citep{GehrelsEtal04}, such as NGC 2617 \citep{ShappeeEtal14}, NGC 5548 \citep{McHardyEtal14,EdelsonEtal15,FausnaughEtal16,StarkeyEtal17}, NGC 4151 \citep{EdelsonEtal17}, MCG+08-11-011 and NGC 2617 \citep{FausnaughEtal18}. 
More recently, other studies have used large quasar samples from various wide field surveys to measure accretion disk sizes: \cite{JiangEtal17} used 39 quasars from Pan-STARRS, \cite{MuddEtal18} had 15 quasars on the supernova fields from the Dark Energy Survey (DES), and \cite{HomayouniEtal18} used 95 quasars from the Sloan Digital Sky Survey (SDSS). 
Most notably \cite{YuEtal18}, presented measurements using high cadence light curves (1 day) in a sample of 23 quasars in the standard star fields and in the supernova C fields of DES.
However, each study found different systematic trends of the measured sizes and prediction of the thin-disk model. 
A plausible explanation might be a bias in the time delay measurements between multi-band light curves when using methods based on interpolated cross-correlation function \citep[ICCF][]{PetersonEtal98} and \JAVELIN\ \citep{ZuEtal11,ZuEtal13}.
Both curve-shifting techniques are able to capture the appropriate lags when the line light curve is a smoothed and shifted version of the continuum \citep{YuEtal19}, but may not be valid when the transfer function of light curves becomes asymmetric. 
Notably, there exist an algorithm that fits directly the reprocessing disk model, named the Continuum REprocessed AGN Markov Chain Monte Carlo code \citep[\CREAM][]{StarkeyEtal16,StarkeyEtal17} that was reported by \citet[][]{FausnaughEtal18} to agree with \JAVELIN's measurement.

In this work, we simulate realistic light curves, as can be obtained with the Large Synoptic Survey Telescope (LSST) in order to test and compare several tools to measure the time lags between light curves taken in multiple photometric bands. 
In addition to ICCF,
 \JAVELIN, and \CREAM, we also use \PyCS\ \citep{TewesEtal13, BonvinEtal16}, a toolbox for time-delay measurements in strongly lensed AGNs developed by the COSMOGRAIL collaboration\footnote{http://www.cosmograil.org}.
 
We present our simulations in \sref{sec:simulation} and our results in \sref{sec:result} along with their implications for future observational strategies to adopt. The conclusions are described in \sref{sec:conclusion}. 
Throughout this paper, we choose $\Lambda$CDM cosmology with $H_0=70$ km $\rm{s}^{-1} \rm{Mpc}^{-1}$, $\Omega_m=0.3$, and $\Omega_\Lambda=0.7$.

\section{Simulating continuum light curves}
\label{sec:simulation}
We introduce here the models for our light-curve simulations and the related asymmetric transfer functions which distort multi-band light curves.
We then use LSST-like light curves to compare disk-size measurements using \JAVELIN, \CREAM, and \PyCS\ and evaluate the biases in these measurements.

\subsection{Thin-disk + lamp-post model}
\label{subsec:model}

We consider a non-relativistic thin-disk model, emitting black-body radiation \citep{Shakura&Sunyaev73}, where the central source is assumed to be a ``lamp post'', located closely above the black hole \citep{CackettEtal07,StarkeyEtal16}.
The time-variable temperature profile of disk can be expressed as 
\be
T^4(R,t)=T^4_0(R)\left[1+f(t-\tlag)\right],
\ee
where $f(t-\tlag)$ is the small fluctuation lagged by the light traveling time $\tlag=(1+z)R/c$. In other words, $f(t)$ is a driving variable source at the center of the AGN.
The unperturbed temperature profile $T_0$ at rest wavelength $\lambda$ can be expressed as 
\be
T_0(R)=\frac{hc}{\lambda k}
\left(\frac{R_\lambda}{R}\right)^{3/4},
\ee
where 
$h$ and $k$ are the Planck and the Boltzmann constants, respectively, and we ignore the inner edge of the disk.
$R_\lambda$ is the radius where the disk temperature matches the photon wavelength ($kT=hc/\lambda$):
\be
R_\lambda
=9.7\times 10^{15} {\rm cm}
\left(\frac{\lambda}{\rm \mu m} \right)^{4/3}
\left(\frac{\MBH}{10^9\Msun} \right)^{2/3}
\left(\frac{L}{\eta L_{\rm E}} \right)^{1/3},
\label{eqn:source_size}
\ee
where $L/L_{\rm E}$ is the luminosity in unit of the Eddington luminosity $L_{\rm E}$ and $\eta$ is the accretion efficiency \citep{MorganEtal10}. 
We assume the accretion disk flux arises from the black-body radiation described by Planck's law
\be
I(R,t)\propto
\frac{1}{\exp(\frac{hc}{\lambda kT})-1}=
\left[\exp(\xi)-1\right]^{-1},
\ee
where
\be
\xi = \left(\frac{R}{R_\lambda}\right)^{3/4} \propto T^{-1}.
\ee
When the temperature variations are small, we have $\delta T \propto f(t-\tlag)/\xi$ and the time-variable surface brightness can be expressed as \citep{Tie&Kochanek18}
\be
\delta I(R,t) = \frac{\partial I}{\partial \xi}\frac{\partial \xi}{\partial T}\delta T 
\propto f(t-\tlag)G_{\lambda}(\xi),
\label{eqn:dI}
\ee
where 
\be
G_{\lambda}(\xi)=\frac{\xi \exp(\xi)}{\left[\exp(\xi)-1\right]^2}. 
\ee
The observed flux is the sum of photons emitted by all points on the disk, the observed photons from the outer disk being responses to irradiation emitted from the central region at earlier times, due to the lamp-post delay $\tlag$.
We can express the fluctuation of the flux as:
\be
\delta F_{\rm obs}(t) = \int\delta I(R,t)dA \propto \int_0^\infty G_{\lambda}(\xi)f(t-\tlag) 2 \pi R dR.
\ee
Therefore, $G_{\lambda}(\xi)$ plays the role of a weight function, and we can now construct a distribution of photon lag, $\delta t$, as a transfer function for an accretion disk:
\be
\label{eqn:transfer}
P_{\delta t}(t) dt= \frac{G_{\lambda}(\xi)R dR}{\int G_{\lambda}(\xi) R dR}.
\ee

\begin{figure*}
\centering
\includegraphics[scale=0.6]{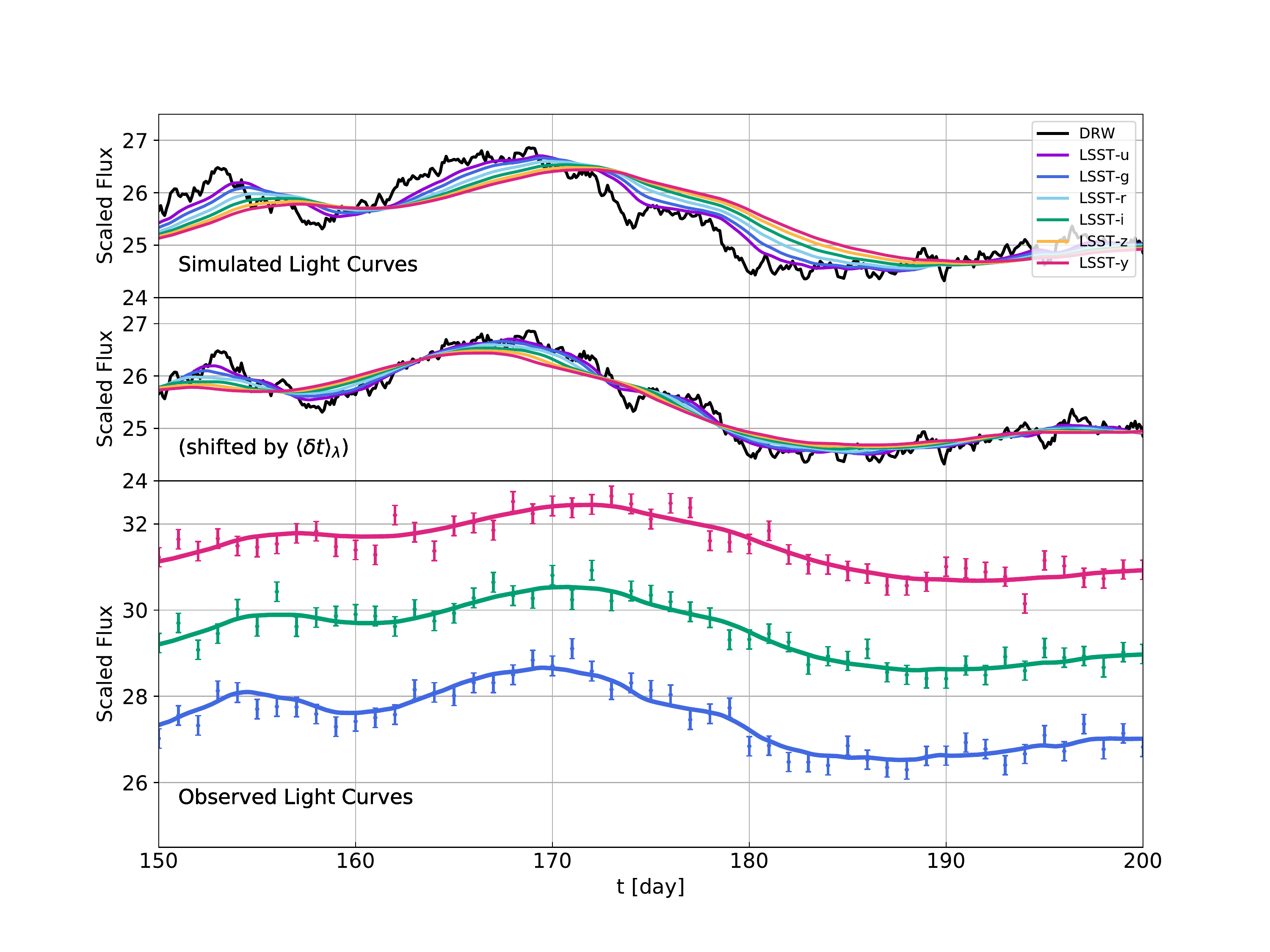}
\caption{
{\it Top:} Simulated light curves for the LSST filters before time sampling.
The black curve is a DRW using $\tau=200$~days and $SF_{\inf}=2$ as a centrally driven source.  
The transfer functions shown on the top panel of \fref{fig:SimGau} are applied to simulate distorted (skewed) multi-band light curves. 
{\it Middle:} The curves are shifted back by their mean delay, $\DTmean$, showing that there are residual misalignments due to convolution by a skewed kernel that may bias delay measurements toward the peak of an asymmetric delay distribution rather than its mean.
{\it Bottom:} Example of the simulated data as may be obtained with the LSST $g$, $r$ and $y$ bands. We simulate the light curves over a period of 1000 days with a 1-day cadence. The error bars are all $\Delta m=0.01$ mag rms.
The color code is the same as in \fref{fig:SimGau}.
}
\label{fig:LCRecov}
\end{figure*}

\subsection{Properties of the simulated light curves}
\label{subsec:light_curve}

To investigate how the source-size measurements are biased due to imperfect time-delay estimates, we design two types of light curves. First a single Gaussian to illustrate how the peak positions shift when convolving with the time-delay transfer function. Second we use light curves represented by a damped random walk (DRW) model, which is currently thought to describe well AGN variability \citep{KellyEtal09,KozlowskiEtal10,ZuEtal13}.

We generate multi-band light curves based on the LSST \ugrizy\ bands (\fref{fig:SimGau}).
In our toy simulation, we chose a quasar redshift of $z=0.5$, a black hole mass $\MBH=2\times10^8~\Msun$, an Eddington ratio of $L/L_{\rm E}=0.1$, and an accretion efficiency of $\eta=0.1$.
Using \eref{eqn:source_size}, the corresponding disk size in the $u$ band is $R_{\LSSTu}= 5.078 \times 10^{14}$~cm $=0.196$~light-day.
Note that when calculating $R_\lambda$ with \eref{eqn:source_size}, we account for the quasar redshift, i.e. $\lambda=\lambda_{\rm obs}/(1+z)$.
The mean delay of each transfer function, given by the thin-disk with lamp-post model can be obtained from:
\be
\DTmean
= \int P_{\delta t} \frac{(1+z)R}{c} dt 
= 5.04 (1+z) \frac{R_\lambda}{c},
\label{eqn:mean_delay}
\ee
which is so-called \textit{the geometric delay}, as it is related to a delayed emission from the different regions of the source \citep{BonvinEtal19}.
We display the transfer function in each band on the top panel of \fref{fig:SimGau}
with $\DTmean$, the mean of these distributions, labelled as vertical dot-dashed lines.
Here we show the effect for a face-on disk, but note that larger inclination angle would introduce even larger bias between the peak and mean values \citep{StarkeyEtal16}.
Furthermore, we ignore the emission lines from the BLR as the latter does not contaminate much the continuum lag measurements \citep[e.g.][]{YuEtal18}.

We first simulate a single Gaussian as a driving-source emission of the lamp-post model, $f(t)$. The black curve on the bottom panel of \fref{fig:SimGau} has been generated using a mean of 0 day and $\sigma=1$ day. The observed flux is the integrated flux of the photons traveling from each region of the disk, with a time lag of $t_{\rm lag}=(1+z)R/c$.

We then construct a transfer function for each filter, and we convolve it with the source emission function, as presented on the bottom panel of \fref{fig:SimGau}. 
The new curves are not only shifted in time, but also smeared asymmetrically, i.e., skewed.
The observed peaks of emission intensity of each curve are indicated as vertical dotted lines. 
In the inset, we highlight with colored wedges the differences between the peak positions and the mean delays with colored wedges.
We note that for a sharper source function (i.e., $\sigma<1$~day), the differences between the peaks and the mean intensities are larger. As a consequence any method measuring time lags from sharp "peaky" structures of similar size to the transfer function will be biased as they are not measuring the mean lag, $\DTmean$, which is the physically relevant quantity.

Next, we create more realistic light curves using a DRW model from {\tt astroML} \citep{astroML,astroMLText} for the driving source emission with a characteristic time-scale $\tau=200$ days at the rest frame and a structure function at infinity $SF_{\inf}=2$ (flux unit), which describes the long-term variability of the light curve.
These DRW parameters are typical of the quasars in the DES sample \citep{YuEtal18}.
Using the same transfer functions as before, we now generate the light curves shown on the top panel of \fref{fig:LCRecov}.
In the middle panel of \fref{fig:LCRecov}, we shifted the curves back by $\DTmean$, to show that the positions, in time, of the minima and maxima are then earlier than in the original DRW light curve.

In practice, light curves are not perfectly continuous. 
To mimic real-life observations, we sample our light curves using photometric noise with a rms scatter of $\Delta m=0.01$ mag over a period of 1000 days with a 1-day cadence. We also remove data falling in season gaps of 120 days every 240 days. In a first experiment, we do not include data loss due to bad weather or technical problem as our purpose is to illustrate measurement biases even with fairly ideal data.
These light curves are presented as the data points on the bottom panels of \fref{fig:LCRecov} and the right panels of \fref{fig:cream}.

We generate 25 DRW realizations of simulated light curves for \PyCS\ considering 25 noise realizations and \JAVELIN/\JAVELINEXT\ considering one noise realization. 
We run \CREAM\ with one of the simulations due to the slow speed computation. This still allows to demonstrate the impact of skewed transfer functions.

\section{Result}
\label{sec:result}

\begin{figure*}
\centering
\includegraphics[width=0.99\textwidth]{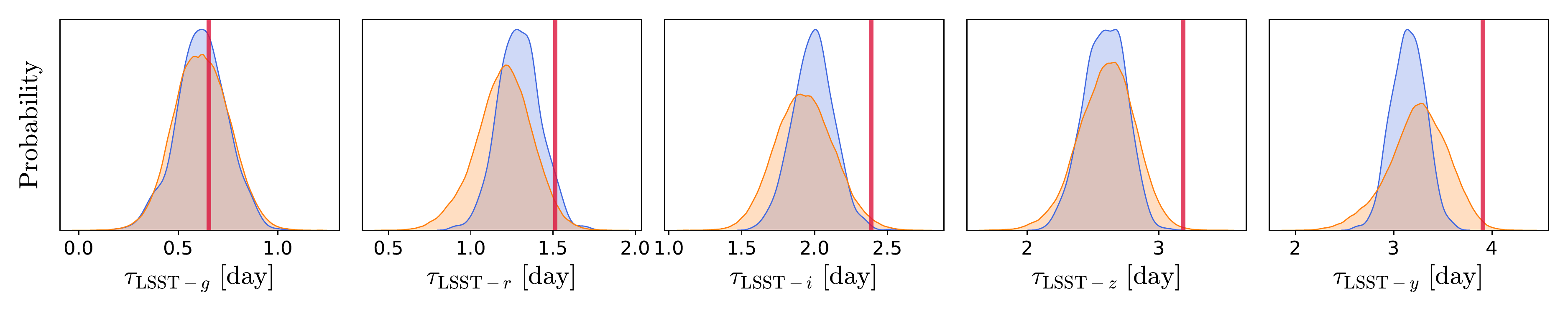}
\begin{minipage}[l]{0.30\textwidth}
\vspace{0.33cm}
\includegraphics[width=0.99\textwidth]{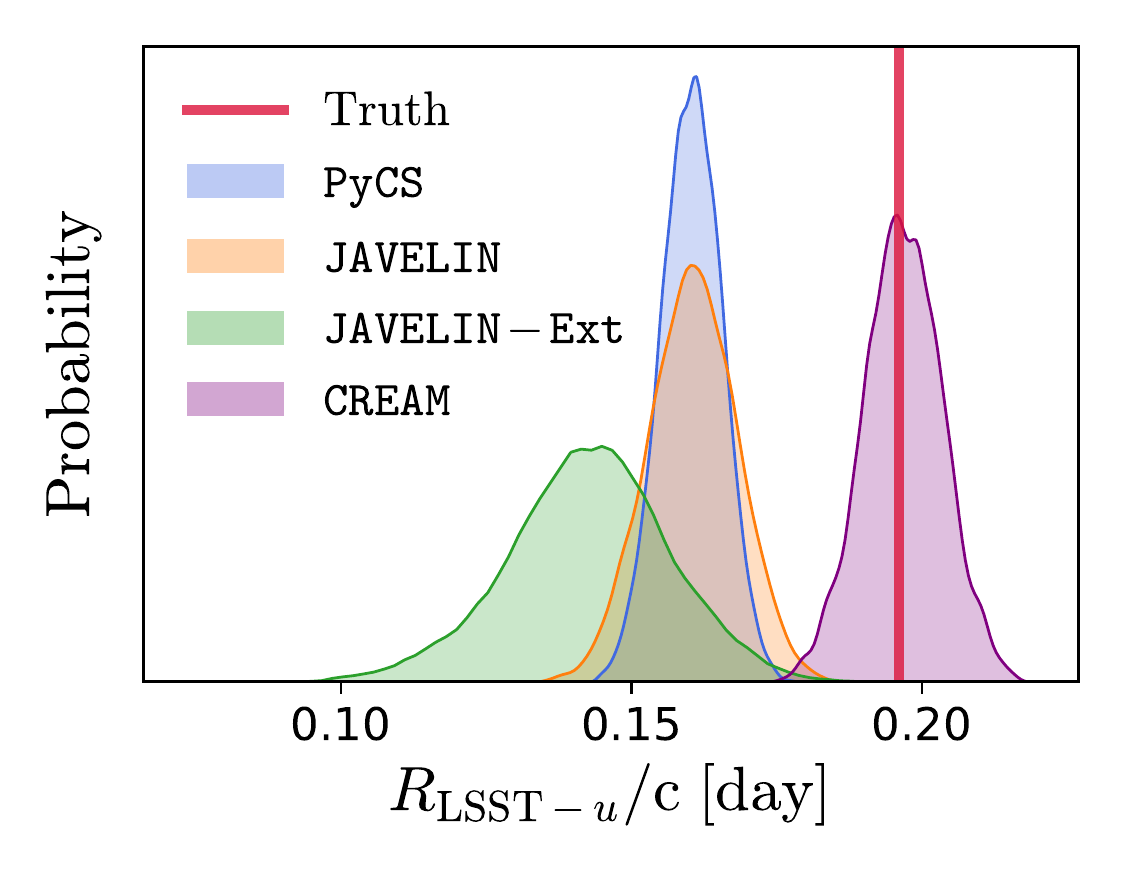}
\end{minipage}
\begin{minipage}[c]{0.69\textwidth}
    \vspace{-0.5cm}
    \centering
    \begin{tabular}{c  c c c c  c}
  & \PyCS & \JAVELIN & \JAVELINEXT & \CREAM & Truth \\ \hline
\vspace{-0.3cm}
& & & & & \\
\vspace{0.1cm} $\tau_{\mathrm{LSST}-g}$& $0.618_{-0.115}^{+0.124}$ & $0.616_{-0.136}^{+0.141}$ & - & - & 0.653 \\ 
\vspace{0.1cm} $\tau_{\mathrm{LSST}-r}$& $1.291_{-0.122}^{+0.130}$ & $1.206_{-0.173}^{+0.162}$ & - & - & 1.514 \\ 
\vspace{0.1cm} $\tau_{\mathrm{LSST}-i}$& $1.995_{-0.140}^{+0.136}$ & $1.929_{-0.196}^{+0.206}$ & - & - & 2.390 \\ 
\vspace{0.1cm} $\tau_{\mathrm{LSST}-z}$& $2.592_{-0.164}^{+0.149}$ & $2.613_{-0.220}^{+0.201}$ & - & - & 3.184 \\ 
\vspace{0.1cm} $\tau_{\mathrm{LSST}-y}$& $3.147_{-0.172}^{+0.172}$ & $3.263_{-0.312}^{+0.287}$ & - & - & 3.908 \\ 

\hline 
\vspace{-0.3cm}
& & & & & \\
 \vspace{0.1cm} $R_{\mathrm{LSST}-u}\mathrm{/c}$ & $0.160_{-0.005}^{+0.005}$ & $0.161_{-0.008}^{+0.008}$ & $0.143_{-0.014}^{+0.013}$ & $0.197_{-0.007}^{+0.007}$ &0.196 \\
    \hline
    \end{tabular}
\end{minipage}
\caption{
Probability distributions of measured time lags $\tau_\LSSTgrizy$ (top-row), relative to the reference band \LSSTu, and the source size $R_{\LSSTu}$ (bottom-left), using \PyCS\ (blue), \JAVELIN\ (orange), \JAVELINEXT\ (green), and \CREAM\ (purple). 
The true delays and sizes are shown as red lines. 
In the table, the values represent the 50th, 16th and 84th percentiles of the respective probability distributions.
The last column (Truth) shows the input value of $\left< \delta t \right>_\LSSTgrizy - \left< \delta t \right>_{\LSSTu}$ using \eref{eqn:mean_delay}.
The results from \PyCS\ and \JAVELIN\ are comparable and are 20\% smaller than the true size.
\JAVELINEXT\ leads to even smaller sizes (30\% smaller) but \CREAM\ is leading to unbiased measurements.
}
\label{fig:delay_measure}
\end{figure*}

The conventional approach to estimate disk sizes is to measure the time delay between light curves observed in different bands and then fit the lags according to a given disk model.
In this work, we choose the $u$ band as the reference band, since its light curve is less distorted, and we employ \PyCS\ and \JAVELIN\ to measure time lags.

\PyCS\ is a publicly available python package developed by the COSMOGRAIL collaboration \citep{TewesEtal13, BonvinEtal16} including two main point-estimators of time delays. One of them is the free-knot splines estimator, that we use here. \PyCS\ does not assume any physical AGN model, and its use is thus completely data driven.

\JAVELIN\ models the variability of AGNs as DRW, assuming light curves at various wavelengths are shifted, scaled, and smoothed versions of the central variability. In practice the shift is done using a convolution of the central DRW by time-shifted top-hat functions \citep{ZuEtal11,ZuEtal13}. 
The width of top-hat of each filter is a free parameter. As a top-hat function is obviously not skewed, we expect a bias in time-lag measurements when applied to continuum light curves.

\JAVELIN\ has been used widely in emission-line and continuum reverberation mapping observations \citep{ShappeeEtal14,FausnaughEtal16,JiangEtal17}.
Due to the difficulty of measuring short delays, \cite{MuddEtal18} have developed an extension of \JAVELIN\ to implement the $\tau\propto\lambda^{4/3}$ scaling of thin-disk models (hereafter \JAVELINEXT), which can directly fit for a disk size $R_{\lambda}$ using all the available photometric light curves simultaneously.

Furthermore, we apply \CREAM\ on our simulated light curves. 
\CREAM\ is designed to fit the more realistic transfer functions of the lamp-post model, for which it infers a posterior probability distribution for the AGN disk size.
In this work, we limit ourselves to face-on disk for both \CREAM\ and \JAVELINEXT. In doing so, we fix the scaling relation of disk size as $R_{\lambda} \propto \lambda^{4/3}$ in accordance to our simulations. We note that in real-life use, these parameters are not necessarily known, and must be marginalized over. 

Finally, we also test the conventional interpolated cross-correlation function (ICCF) method \citep{PetersonEtal98,SunEtal18}. This leads to the same conclusion as \cite{JiangEtal17,YuEtal18}, i.e. the lag distributions from ICCF are significantly wider ($\gtrsim 0.8$~day) than the distributions from the previous three methods, to a point that statistical error vastly dominate the systematic errors we wish to study here. 
We therefore do not consider further the ICCF method.

In the top row of \fref{fig:delay_measure}, we display the probability distributions of the measured time lags $\tau_\lambda$ relative to the reference band ($\lambda_0=\LSSTu$ in this work) from \PyCS\ and \JAVELIN. 
We indicate the true delays as red vertical lines, i.e., $\left< \delta t \right>_\lambda - \left< \delta t \right>_{\lambda_0}$ using \eref{eqn:mean_delay}.
To estimate the disk size, we adopt least-square fitting using the measured delays $\tau_{\lambda}$ between the LSST bands.
This is
\be
\chi^2 = \sum_{\lambda}\frac{\left[ \Delta t_{\lambda}-\left<\delta t\right>_{\lambda} \right]^2}{\sigma_{\lambda}^2},
\ee
where 
\be
\Delta t_{\lambda}-\left<\delta t\right>_{\lambda} = \tau_{\lambda}+
5.04(1+z)\frac{R_{\lambda_0}}{c}
[1-(\lambda/\lambda_0)^{4/3}]
\ee
and where $\sigma_{\lambda}$ are the standard deviations of the distributions. The probability distributions for the source size at the reference band $R_\LSSTu$ are shown on the bottom-left panel of \fref{fig:delay_measure}.
Evidently, the smaller size measurements of \PyCS\ and \JAVELIN\ are coming from the underestimation of time lags in comparison with the mean delays of the transfer function $\left< \delta t \right>_\lambda - \left< \delta t \right>_{\lambda_0}$. The effect is stronger for light curves displaying sharp peaks, i.e. when the driving light curve is dominated by high frequency structures acting on time-scales similar to the (temporal) size of the transfer function. Both \PyCS\ and \JAVELIN\ seem sensitive to such structures, which are most affected by the skewed transfer functions. This results in estimated source sizes about 20\% smaller than the true size for our simulations, but this should degrade even more with increasing level of sharp structures in the DRW, hence the bias depends on every object.

In contrast to the previous methods, both \CREAM\ and \JAVELINEXT\ fit the source size directly, but the main differences are: 1- \JAVELINEXT\ adopts a shifted top-hat transfer function while \CREAM\ has a realistic thin-disk model, and 2- \CREAM\ represents the driving source with a prior on the Fourier power density spectrum while \JAVELINEXT\ assumes that the driving light curve is a DRW.
The posteriors of DRW parameters from \JAVELIN\ are shown in \fref{fig:javelin_drw}.

\begin{figure*}[t!]
\centering
\includegraphics[scale=0.85]{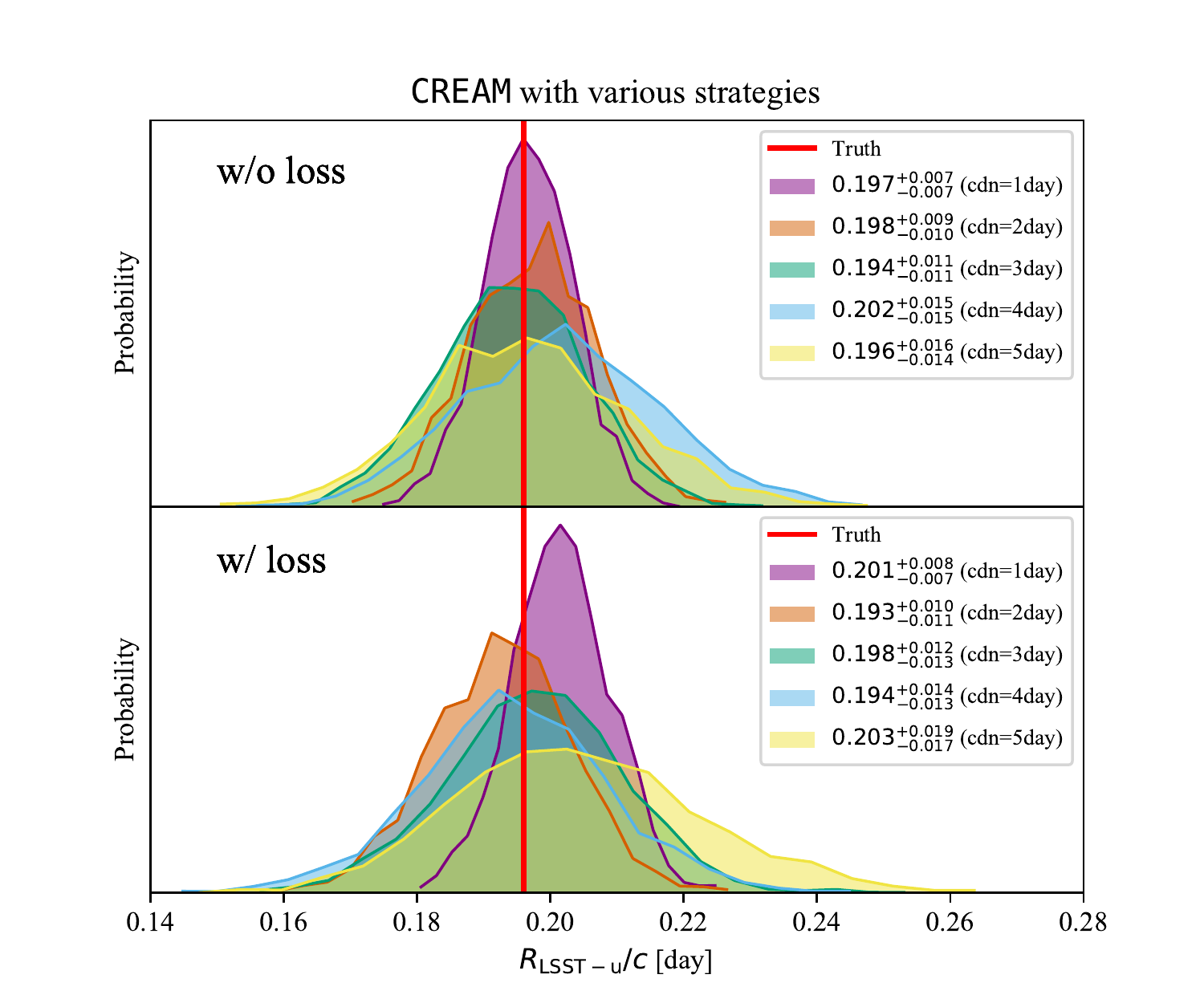}
\caption{
Probability distributions of the source size $R_{\LSSTu}$ using \CREAM.
The simulated light curves are the same as shown on the right panels of \fref{fig:cream}, and we increase cadence (cdn) from 1 to 5 days ({\it Top}) and further add inter-night gaps and 10\% loss of time ({\it Bottom}).
\CREAM\ measures the disk size with an uncertainty below 10\% even when a cadence is 5 days with data loss. 
}
\label{fig:cdn_loss}
\end{figure*}

\subsection{Consequences on future observational strategies}
\label{sec:cdn_loss}

\CREAM\ outputs the posterior of accretion rate $\dot{M}$ and the mean delay of each transfer function $\DTmean$.
The number of total iteration is 20,000, and only the samples after iteration larger 5,000 are used for size measurement.
To avoid confusion in converting \CREAM's measurement of $\dot{M}$ to $R_\lambda$, we use $\DTmeanLSSTu$ as our estimate to measure the disk size, using \eref{eqn:source_size}.
The MCMC fit of \CREAM\ is illustrated in \fref{fig:cream}.
More inputs/outputs of \CREAM\ are discussed in \aref{appendix}.

The results of \JAVELINEXT\ and \CREAM\ are also shown in \fref{fig:delay_measure}. We note an even larger bias for \JAVELINEXT\ which lead to an underestimate of the size by 30\%. The result of \JAVELINEXT\ has the same trend as shown in Figure 11 of \cite{MuddEtal18}.
\CREAM\ obtains the unbiased measurement with a rms uncertainty of $\sim4\%$.
However, there are so far very few direct comparisons of different time lag measurement algorithms on real data.
Based on the present experiment, we conclude that it is crucial to consider asymmetric transfer functions to infer the disk size using continuum reverberation mapping data.

We now investigate the precision of the disk size measurement from simple observational strategies using \CREAM, as this is the method giving the best performances in the previous sections.
We use the same light curve simulation as for \CREAM, as shown in \fref{fig:cream} ($\Delta m = 0.01$), but we increase cadence gradually from 1 day to 5 days. 
Furthermore, we randomly add 3 inter-night gaps to each observation season (240 days) and $10\%$ loss of observation to mimic bad weather or technical issues at the telescope.
Each inter-night gap is one week. These observing conditions are typical for the COSMOGRAIL program and should also apply to e.g. LSST.
The numbers of epochs of each observation strategy is listed in \tref{tab:strategy}.
The total number of iterations is 20,000. 
We choose the samples after iteration 5,000 and then measure the probability distributions for $\DTmeanLSSTu$.
The corresponding source sizes at the reference band $R_{\LSSTu}$ from \CREAM\ are shown in \fref{fig:cdn_loss}.
In general, \CREAM\ can measure disk sizes with a rms error below 10\%, even for a cadence of 5 days and with data loss.
A cadence better than 2 days even achieves errors below the 5\% level.

\begin{table*}[]
    \centering
    \resizebox{\textwidth}{!}{
    \begin{tabular}{|c|c|c|c|c|c|c|c|c|c|c|}
    \hline
    cadence & \multicolumn{2}{c|}{1} & \multicolumn{2}{c|}{2} &
    \multicolumn{2}{c|}{3} & \multicolumn{2}{c|}{4} & \multicolumn{2}{c|}{5} \\
    \hline
    loss & no & yes & no & yes & no & yes & no & yes & no & yes \\
    \hline
    \hline
	epochs & 719 & 602 & 359 & 299 & 239 & 195 & 179 & 147 & 143 & 122 \\
	\hline
	$\left< \delta t \right>_{\rm LSST-u}$ & ${1.487}^{+0.051}_{-0.049}$ & ${1.522}^{+0.059}_{-0.053}$ & ${1.494}^{+0.068}_{-0.075}$ & ${1.458}^{+0.077}_{-0.080}$ & ${1.468}^{+0.081}_{-0.084}$ & ${1.497}^{+0.094}_{-0.098}$ & ${1.524}^{+0.113}_{-0.114}$ & ${1.468}^{+0.104}_{-0.100}$ & ${1.481}^{+0.120}_{-0.107}$ & ${1.536}^{+0.143}_{-0.131}$ \\
	\hline
    \end{tabular}
    }
    \vskip 5pt
    \caption{
    Number of epochs for each observation strategy using \CREAM.
    The total observation time is 1000 days with a season gap of 120 days every 360 days. 
    For the simulations with data loss, we randomly add 3 inter-night gaps to each observation season (240 days) and 10\% loss of time. 
    We list the 50th, 16th and 84th percentiles of the $\DTmeanLSSTu$ probability distributions of the MCMC samples from iterations 5,000 to 20,000.
    The corresponding disk measurements are shown in \fref{fig:cdn_loss}.
    }
    \label{tab:strategy}
\end{table*}

\section{Conclusions}
\label{sec:conclusion}

We show that most current methods to analyze continuum reverberation mapping data lead to underestimates of accretion-disk sizes if the transfer function is skewed.

To illustrate our finding, we simulate light curves using an AGN model based on a combination of thin-disk, lamp-post and damped random walk for the driving function. We estimate the time lags using both \PyCS's free-knot splines estimator and \JAVELIN\, and we use the measured lags to derive the disk size. We also use \JAVELINEXT\ and \CREAM\ to directly fit the disk size, with the former adopting a shifted top-hat transfer function.
Our findings are as follows:
\begin{itemize}
    \item Because the transfer function of thin-disk model is asymmetric, multi-band continuum light curves are not only shifted and smoothed w.r.t. each other, but also skewed.
    \item Curve-shifting techniques that are sensitive to sharp features, acting on similar time scales as the transfer function, underestimate multi-band time delays by up to 20\%, hence translating into a source size estimate also 20\% smaller than the truth. We note that the number quoted here depends on the level of high frequency structures (e.g. sharp peaks) in the actual DRW used.
    \item Direct disk size estimates using \JAVELINEXT\ do not perform better, with fitted size being 30\% smaller than the truth.
    \item Taking the proper transfer functions into account, such as \CREAM, is essential to reach an unbiased measurement of the disk size.
    \item To achieve the size measurement with errors below 5\%, a cadence of at least one observation every 2 days is needed, assuming photometric errors of the order of $\Delta m=0.01$ mag rms over a period of $\sim 3$ years.
\end{itemize}
A long-standing problem in quasar accretion disk studies is that measurements of their size are larger than predictions of the thin-disk model by factors as large as $2-3$.
Recently, \cite{YuEtal18} reported that the measured disk sizes are consistent with the predictions after taking the disk variability into account, assuming that disks illuminate following the lamp-post model, because this increases the flux-weighted mean disk size by up to 50\% \cite[See Equations (8) and (9) in][]{YuEtal18}. 
However, the discrepancy still exists if the effect of the skewed transfer function is ignored.
In this work, we choose the traditional disk model to demonstrate the effect.
Although the details of the transfer function may vary from other models, we show that accounting for its skewness, which is a general property shared by many models, is necessary to converge towards unbiased measurements of the disk size. 

Our results based on numerical experiments suggest that future generations of continuum reverberation mapping studies should consider the transfer function shape in detail, or potentially attempt to reconstruct it during the time-lag measurement process.


\section*{Acknowledgements}

This research is supported by the Swiss National Science
Foundation (SNSF) and by the European Research Council (ERC) under the European Union’s Horizon 2020 research and innovation program (COSMICLENS: grant agreement No 787866). 
We thank Dominique Sluse for useful discussion.
We would also like to thank the anonymous referee for the constructive comments on this work.


\bibliographystyle{aa}
\bibliography{reference}

\begin{thebibliography}{33}
\expandafter\ifx\csname natexlab\endcsname\relax\def\natexlab#1{#1}\fi

\bibitem[{{Blandford} \& {McKee}(1982)}]{Blandford&Mckee82}
{Blandford}, R.~D. \& {McKee}, C.~F. 1982, \apj, 255, 419

\bibitem[{{Bonvin} {et~al.}(2016){Bonvin}, {Tewes}, {Courbin}, {Kuntzer},
  {Sluse}, \& {Meylan}}]{BonvinEtal16}
{Bonvin}, V., {Tewes}, M., {Courbin}, F., {et~al.} 2016, \aap, 585, A88

\bibitem[{{Bonvin} {et~al.}(2019){Bonvin}, {Tihhonova}, {Millon}, {Chan},
  {Savary}, {Huber}, \& {Courbin}}]{BonvinEtal19}
{Bonvin}, V., {Tihhonova}, O., {Millon}, M., {et~al.} 2019, \aap, 621, A55

\bibitem[{{Cackett} {et~al.}(2007){Cackett}, {Horne}, \&
  {Winkler}}]{CackettEtal07}
{Cackett}, E.~M., {Horne}, K., \& {Winkler}, H. 2007, \mnras, 380, 669

\bibitem[{{Edelson} {et~al.}(2017){Edelson}, {Gelbord}, {Cackett}, {Connolly},
  {Done}, {Fausnaugh}, {Gardner}, {Gehrels}, {Goad}, {Horne}, {McHardy},
  {Peterson}, {Vaughan}, {Vestergaard}, {Breeveld}, {Barth}, {Bentz},
  {Bottorff}, {Brandt}, {Crawford}, {Dalla Bont{\`a}}, {Emmanoulopoulos},
  {Evans}, {Figuera Jaimes}, {Filippenko}, {Ferland}, {Grupe}, {Joner},
  {Kennea}, {Korista}, {Krimm}, {Kriss}, {Leonard}, {Mathur}, {Netzer},
  {Nousek}, {Page}, {Romero-Colmenero}, {Siegel}, {Starkey}, {Treu}, {Vogler},
  {Winkler}, \& {Zheng}}]{EdelsonEtal17}
{Edelson}, R., {Gelbord}, J., {Cackett}, E., {et~al.} 2017, \apj, 840, 41

\bibitem[{{Edelson} {et~al.}(2015){Edelson}, {Gelbord}, {Horne}, {McHardy},
  {Peterson}, {Ar{\'e}valo}, {Breeveld}, {De Rosa}, {Evans}, {Goad}, {Kriss},
  {Brandt}, {Gehrels}, {Grupe}, {Kennea}, {Kochanek}, {Nousek}, {Papadakis},
  {Siegel}, {Starkey}, {Uttley}, {Vaughan}, {Young}, {Barth}, {Bentz},
  {Brewer}, {Crenshaw}, {Dalla Bont{\`a}}, {De Lorenzo-C{\'a}ceres}, {Denney},
  {Dietrich}, {Ely}, {Fausnaugh}, {Grier}, {Hall}, {Kaastra}, {Kelly},
  {Korista}, {Lira}, {Mathur}, {Netzer}, {Pancoast}, {Pei}, {Pogge},
  {Schimoia}, {Treu}, {Vestergaard}, {Villforth}, {Yan}, \&
  {Zu}}]{EdelsonEtal15}
{Edelson}, R., {Gelbord}, J.~M., {Horne}, K., {et~al.} 2015, \apj, 806, 129

\bibitem[{{Fausnaugh} {et~al.}(2016){Fausnaugh}, {Denney}, {Barth}, {Bentz},
  {Bottorff}, {Carini}, {Croxall}, {De Rosa}, {Goad}, {Horne}, {Joner},
  {Kaspi}, {Kim}, {Klimanov}, {Kochanek}, {Leonard}, {Netzer}, {Peterson},
  {Schn{\"u}lle}, {Sergeev}, {Vestergaard}, {Zheng}, {Zu}, {Anderson},
  {Ar{\'e}valo}, {Bazhaw}, {Borman}, {Boroson}, {Brandt}, {Breeveld}, {Brewer},
  {Cackett}, {Crenshaw}, {Dalla Bont{\`a}}, {De Lorenzo-C{\'a}ceres},
  {Dietrich}, {Edelson}, {Efimova}, {Ely}, {Evans}, {Filippenko}, {Flatland},
  {Gehrels}, {Geier}, {Gelbord}, {Gonzalez}, {Gorjian}, {Grier}, {Grupe},
  {Hall}, {Hicks}, {Horenstein}, {Hutchison}, {Im}, {Jensen}, {Jones},
  {Kaastra}, {Kelly}, {Kennea}, {Kim}, {Korista}, {Kriss}, {Lee}, {Lira},
  {MacInnis}, {Manne-Nicholas}, {Mathur}, {McHardy}, {Montouri}, {Musso},
  {Nazarov}, {Norris}, {Nousek}, {Okhmat}, {Pancoast}, {Papadakis}, {Parks},
  {Pei}, {Pogge}, {Pott}, {Rafter}, {Rix}, {Saylor}, {Schimoia}, {Siegel},
  {Spencer}, {Starkey}, {Sung}, {Teems}, {Treu}, {Turner}, {Uttley},
  {Villforth}, {Weiss}, {Woo}, {Yan}, \& {Young}}]{FausnaughEtal16}
{Fausnaugh}, M.~M., {Denney}, K.~D., {Barth}, A.~J., {et~al.} 2016, \apj, 821,
  56

\bibitem[{{Fausnaugh} {et~al.}(2018){Fausnaugh}, {Starkey}, {Horne},
  {Kochanek}, {Peterson}, {Bentz}, {Denney}, {Grier}, {Grupe}, {Pogge}, {De
  Rosa}, {Adams}, {Barth}, {Beatty}, {Bhattacharjee}, {Borman}, {Boroson},
  {Bottorff}, {Brown}, {Brown}, {Brotherton}, {Coker}, {Crawford}, {Croxall},
  {Eftekharzadeh}, {Eracleous}, {Joner}, {Henderson}, {Holoien}, {Hutchison},
  {Kaspi}, {Kim}, {King}, {Li}, {Lochhaas}, {Ma}, {MacInnis}, {Manne-Nicholas},
  {Mason}, {Montuori}, {Mosquera}, {Mudd}, {Musso}, {Nazarov}, {Nguyen},
  {Okhmat}, {Onken}, {Ou- Yang}, {Pancoast}, {Pei}, {Penny}, {Poleski},
  {Rafter}, {Romero- Colmenero}, {Runnoe}, {Sand}, {Schimoia}, {Sergeev},
  {Shappee}, {Simonian}, {Somers}, {Spencer}, {Stevens}, {Tayar}, {Treu},
  {Valenti}, {Van Saders}, {Villanueva}, {Villforth}, {Weiss}, {Winkler}, \&
  {Zhu}}]{FausnaughEtal18}
{Fausnaugh}, M.~M., {Starkey}, D.~A., {Horne}, K., {et~al.} 2018, \apj, 854,
  107

\bibitem[{{Gehrels} {et~al.}(2004){Gehrels}, {Chincarini}, {Giommi}, {Mason},
  {Nousek}, {Wells}, {White}, {Barthelmy}, {Burrows}, {Cominsky}, {Hurley},
  {Marshall}, {M{\'e}sz{\'a}ros}, {Roming}, {Angelini}, {Barbier}, {Belloni},
  {Campana}, {Caraveo}, {Chester}, {Citterio}, {Cline}, {Cropper}, {Cummings},
  {Dean}, {Feigelson}, {Fenimore}, {Frail}, {Fruchter}, {Garmire}, {Gendreau},
  {Ghisellini}, {Greiner}, {Hill}, {Hunsberger}, {Krimm}, {Kulkarni}, {Kumar},
  {Lebrun}, {Lloyd- Ronning}, {Markwardt}, {Mattson}, {Mushotzky}, {Norris},
  {Osborne}, {Paczynski}, {Palmer}, {Park}, {Parsons}, {Paul}, {Rees},
  {Reynolds}, {Rhoads}, {Sasseen}, {Schaefer}, {Short}, {Smale}, {Smith},
  {Stella}, {Tagliaferri}, {Takahashi}, {Tashiro}, {Townsley}, {Tueller},
  {Turner}, {Vietri}, {Voges}, {Ward}, {Willingale}, {Zerbi}, \&
  {Zhang}}]{GehrelsEtal04}
{Gehrels}, N., {Chincarini}, G., {Giommi}, P., {et~al.} 2004, \apj, 611, 1005

\bibitem[{{Homayouni} {et~al.}(2018){Homayouni}, {Trump}, {Grier}, {Shen},
  {Starkey}, {Brandt}, {Hall}, {Horne}, {Kinemuchi}, {I-Hsiu Li}, {McGreer},
  {Sun}, {Ho}, \& {Schneider}}]{HomayouniEtal18}
{Homayouni}, Y., {Trump}, J.~R., {Grier}, C.~J., {et~al.} 2018, ArXiv e-prints
  [\eprint[arXiv]{1806.08360}]

\bibitem[{{Ivezi{\'c}} {et~al.}(2014){Ivezi{\'c}}, {Connolly}, {Vanderplas}, \&
  {Gray}}]{astroMLText}
{Ivezi{\'c}}, {\v Z}., {Connolly}, A., {Vanderplas}, J., \& {Gray}, A. 2014,
  Statistics, Data Mining and Machine Learning in Astronomy (Princeton
  University Press)

\bibitem[{{Jiang} {et~al.}(2017){Jiang}, {Green}, {Greene}, {Morganson},
  {Shen}, {Pancoast}, {MacLeod}, {Anderson}, {Brandt}, {Grier}, {Rix}, {Ruan},
  {Protopapas}, {Scott}, {Burgett}, {Hodapp}, {Huber}, {Kaiser}, {Kudritzki},
  {Magnier}, {Metcalfe}, {Tonry}, {Wainscoat}, \& {Waters}}]{JiangEtal17}
{Jiang}, Y.-F., {Green}, P.~J., {Greene}, J.~E., {et~al.} 2017, \apj, 836, 186

\bibitem[{{Kelly} {et~al.}(2009){Kelly}, {Bechtold}, \&
  {Siemiginowska}}]{KellyEtal09}
{Kelly}, B.~C., {Bechtold}, J., \& {Siemiginowska}, A. 2009, \apj, 698, 895

\bibitem[{{Kochanek}(2004)}]{Kochanek04}
{Kochanek}, C.~S. 2004, \apj, 605, 58

\bibitem[{{Koz{\l}owski} {et~al.}(2010){Koz{\l}owski}, {Kochanek}, {Udalski},
  {Wyrzykowski}, {Soszy{\'n}ski}, {Szyma{\'n}ski}, {Kubiak}, {Pietrzy{\'n}ski},
  {Szewczyk}, {Ulaczyk}, {Poleski}, \& {OGLE Collaboration}}]{KozlowskiEtal10}
{Koz{\l}owski}, S., {Kochanek}, C.~S., {Udalski}, A., {et~al.} 2010, \apj, 708,
  927

\bibitem[{{McHardy} {et~al.}(2014){McHardy}, {Cameron}, {Dwelly}, {Connolly},
  {Lira}, {Emmanoulopoulos}, {Gelbord}, {Breedt}, {Arevalo}, \&
  {Uttley}}]{McHardyEtal14}
{McHardy}, I.~M., {Cameron}, D.~T., {Dwelly}, T., {et~al.} 2014, \mnras, 444,
  1469

\bibitem[{{Morgan} {et~al.}(2018){Morgan}, {Hyer}, {Bonvin}, {Mosquera},
  {Cornachione}, {Courbin}, {Kochanek}, \& {Falco}}]{MorganEtal18}
{Morgan}, C.~W., {Hyer}, G.~E., {Bonvin}, V., {et~al.} 2018, \apj, 869, 106

\bibitem[{{Morgan} {et~al.}(2010){Morgan}, {Kochanek}, {Morgan}, \&
  {Falco}}]{MorganEtal10}
{Morgan}, C.~W., {Kochanek}, C.~S., {Morgan}, N.~D., \& {Falco}, E.~E. 2010,
  \apj, 712, 1129

\bibitem[{{Mudd} {et~al.}(2018){Mudd}, {Martini}, {Zu}, {Kochanek}, {Peterson},
  {Kessler}, {Davis}, {Hoormann}, {King}, {Lidman}, {Sommer}, {Tucker},
  {Asorey}, {Hinton}, {Glazebrook}, {Kuehn}, {Lewis}, {Macaulay}, {Moeller},
  {O'Neill}, {Zhang}, {Abbott}, {Abdalla}, {Allam}, {Banerji},
  {Benoit-L{\'e}vy}, {Bertin}, {Brooks}, {Carnero Rosell}, {Carollo}, {Carrasco
  Kind}, {Carretero}, {Cunha}, {D'Andrea}, {da Costa}, {Davis}, {Desai},
  {Doel}, {Fosalba}, {Garc{\'{\i}}a-Bellido}, {Gaztanaga}, {Gerdes}, {Gruen},
  {Gruendl}, {Gschwend}, {Gutierrez}, {Hartley}, {Honscheid}, {James},
  {Kuhlmann}, {Kuropatkin}, {Lima}, {Maia}, {Marshall}, {McMahon}, {Menanteau},
  {Miquel}, {Plazas}, {Romer}, {Sanchez}, {Schindler}, {Schubnell}, {Smith},
  {Smith}, {Soares-Santos}, {Sobreira}, {Suchyta}, {Swanson}, {Tarle},
  {Thomas}, {Tucker}, {Walker}, \& {DES Collaboration}}]{MuddEtal18}
{Mudd}, D., {Martini}, P., {Zu}, Y., {et~al.} 2018, \apj, 862, 123

\bibitem[{{Peterson} {et~al.}(1998){Peterson}, {Wanders}, {Horne}, {Collier},
  {Alexander}, {Kaspi}, \& {Maoz}}]{PetersonEtal98}
{Peterson}, B.~M., {Wanders}, I., {Horne}, K., {et~al.} 1998, Publications of
  the Astronomical Society of the Pacific, 110, 660

\bibitem[{{Schechter} \& {Wambsganss}(2002)}]{Schechter&Wambsganss02}
{Schechter}, P.~L. \& {Wambsganss}, J. 2002, \apj, 580, 685

\bibitem[{{Shakura} \& {Sunyaev}(1973)}]{Shakura&Sunyaev73}
{Shakura}, N.~I. \& {Sunyaev}, R.~A. 1973, \aap, 24, 337

\bibitem[{{Shappee} {et~al.}(2014){Shappee}, {Prieto}, {Grupe}, {Kochanek},
  {Stanek}, {De Rosa}, {Mathur}, {Zu}, {Peterson}, {Pogge}, {Komossa}, {Im},
  {Jencson}, {Holoien}, {Basu}, {Beacom}, {Szczygie{\l}}, {Brimacombe},
  {Adams}, {Campillay}, {Choi}, {Contreras}, {Dietrich}, {Dubberley},
  {Elphick}, {Foale}, {Giustini}, {Gonzalez}, {Hawkins}, {Howell}, {Hsiao},
  {Koss}, {Leighly}, {Morrell}, {Mudd}, {Mullins}, {Nugent}, {Parrent},
  {Phillips}, {Pojmanski}, {Rosing}, {Ross}, {Sand}, {Terndrup}, {Valenti},
  {Walker}, \& {Yoon}}]{ShappeeEtal14}
{Shappee}, B.~J., {Prieto}, J.~L., {Grupe}, D., {et~al.} 2014, \apj, 788, 48

\bibitem[{{Starkey} {et~al.}(2017){Starkey}, {Horne}, {Fausnaugh}, {Peterson},
  {Bentz}, {Kochanek}, {Denney}, {Edelson}, {Goad}, {De Rosa}, {Anderson},
  {Ar{\'e}valo}, {Barth}, {Bazhaw}, {Borman}, {Boroson}, {Bottorff}, {Brandt},
  {Breeveld}, {Cackett}, {Carini}, {Croxall}, {Crenshaw}, {Dalla Bont{\`a}},
  {De Lorenzo-C{\'a}ceres}, {Dietrich}, {Efimova}, {Ely}, {Evans},
  {Filippenko}, {Flatland}, {Gehrels}, {Geier}, {Gelbord}, {Gonzalez},
  {Gorjian}, {Grier}, {Grupe}, {Hall}, {Hicks}, {Horenstein}, {Hutchison},
  {Im}, {Jensen}, {Joner}, {Jones}, {Kaastra}, {Kaspi}, {Kelly}, {Kennea},
  {Kim}, {Kim}, {Klimanov}, {Korista}, {Kriss}, {Lee}, {Leonard}, {Lira},
  {MacInnis}, {Manne-Nicholas}, {Mathur}, {McHardy}, {Montouri}, {Musso},
  {Nazarov}, {Norris}, {Nousek}, {Okhmat}, {Pancoast}, {Parks}, {Pei}, {Pogge},
  {Pott}, {Rafter}, {Rix}, {Saylor}, {Schimoia}, {Schn{\"u}lle}, {Sergeev},
  {Siegel}, {Spencer}, {Sung}, {Teems}, {Turner}, {Uttley}, {Vestergaard},
  {Villforth}, {Weiss}, {Woo}, {Yan}, {Young}, {Zheng}, \&
  {Zu}}]{StarkeyEtal17}
{Starkey}, D., {Horne}, K., {Fausnaugh}, M.~M., {et~al.} 2017, \apj, 835, 65

\bibitem[{{Starkey} {et~al.}(2016){Starkey}, {Horne}, \&
  {Villforth}}]{StarkeyEtal16}
{Starkey}, D.~A., {Horne}, K., \& {Villforth}, C. 2016, \mnras, 456, 1960

\bibitem[{{Sun} {et~al.}(2018){Sun}, {Grier}, \& {Peterson}}]{SunEtal18}
{Sun}, M., {Grier}, C.~J., \& {Peterson}, B.~M. 2018, {PyCCF: Python Cross
  Correlation Function for reverberation mapping studies}, Astrophysics Source
  Code Library

\bibitem[{{Tewes} {et~al.}(2013){Tewes}, {Courbin}, \& {Meylan}}]{TewesEtal13}
{Tewes}, M., {Courbin}, F., \& {Meylan}, G. 2013, \aap, 553, A120

\bibitem[{{Tie} \& {Kochanek}(2018)}]{Tie&Kochanek18}
{Tie}, S.~S. \& {Kochanek}, C.~S. 2018, \mnras, 473, 80

\bibitem[{{Vanderplas} {et~al.}(2012){Vanderplas}, {Connolly}, {Ivezi{\'c}}, \&
  {Gray}}]{astroML}
{Vanderplas}, J., {Connolly}, A., {Ivezi{\'c}}, {\v Z}., \& {Gray}, A. 2012, in
  Conference on Intelligent Data Understanding (CIDU), 47 --54

\bibitem[{{Yu} {et~al.}(2019){Yu}, {Kochanek}, {Peterson}, {Zu}, {Brandt},
  {Cackett}, {Fausnaugh}, \& {McHardy}}]{YuEtal19}
{Yu}, Z., {Kochanek}, C.~S., {Peterson}, B.~M., {et~al.} 2019, arXiv e-prints,
  arXiv:1909.03072

\bibitem[{{Yu} {et~al.}(2018){Yu}, {Martini}, {Davis}, {Gruendl}, {Hoormann},
  {Kochanek}, {Lidman}, {Mudd}, {Peterson}, {Wester}, {Allam}, {Annis},
  {Asorey}, {Avila}, {Banerji}, {Bertin}, {Brooks}, {Buckley-Geer}, {Calcino},
  {Carnero Rosell}, {Carollo}, {Carrasco Kind}, {Carretero}, {Cunha},
  {D'Andrea}, {da Costa}, {De Vicente}, {Desai}, {Diehl}, {Doel}, {Eifler},
  {Flaugher}, {Fosalba}, {Frieman}, {Garc{\'{\i}}a-Bellido}, {Gaztanaga},
  {Glazebrook}, {Gruen}, {Gschwend}, {Gutierrez}, {Hartley}, {Hinton},
  {Hollowood}, {Honscheid}, {Hoyle}, {James}, {Kim}, {Krause}, {Kuehn},
  {Kuropatkin}, {Lewis}, {Lima}, {Macaulay}, {Maia}, {Marshall}, {Menanteau},
  {Miquel}, {M{\"o}ller}, {Plazas}, {Romer}, {Sanchez}, {Scarpine},
  {Schubnell}, {Serrano}, {Smith}, {Smith}, {Soares-Santos}, {Sobreira},
  {Suchyta}, {Swann}, {Swanson}, {Tarle}, {Tucker}, {Tucker}, \&
  {Vikram}}]{YuEtal18}
{Yu}, Z., {Martini}, P., {Davis}, T.~M., {et~al.} 2018, ArXiv e-prints
  [\eprint[arXiv]{1811.03638}]

\bibitem[{{Zu} {et~al.}(2013){Zu}, {Kochanek}, {Koz{\l}owski}, \&
  {Udalski}}]{ZuEtal13}
{Zu}, Y., {Kochanek}, C.~S., {Koz{\l}owski}, S., \& {Udalski}, A. 2013, \apj,
  765, 106

\bibitem[{{Zu} {et~al.}(2011){Zu}, {Kochanek}, \& {Peterson}}]{ZuEtal11}
{Zu}, Y., {Kochanek}, C.~S., \& {Peterson}, B.~M. 2011, \apj, 735, 80

\end{thebibliography}

\appendix
\section{The fits from \JAVELIN\ and \CREAM}
\label{appendix}

\begin{figure*}
\centering
\includegraphics[scale=0.6]{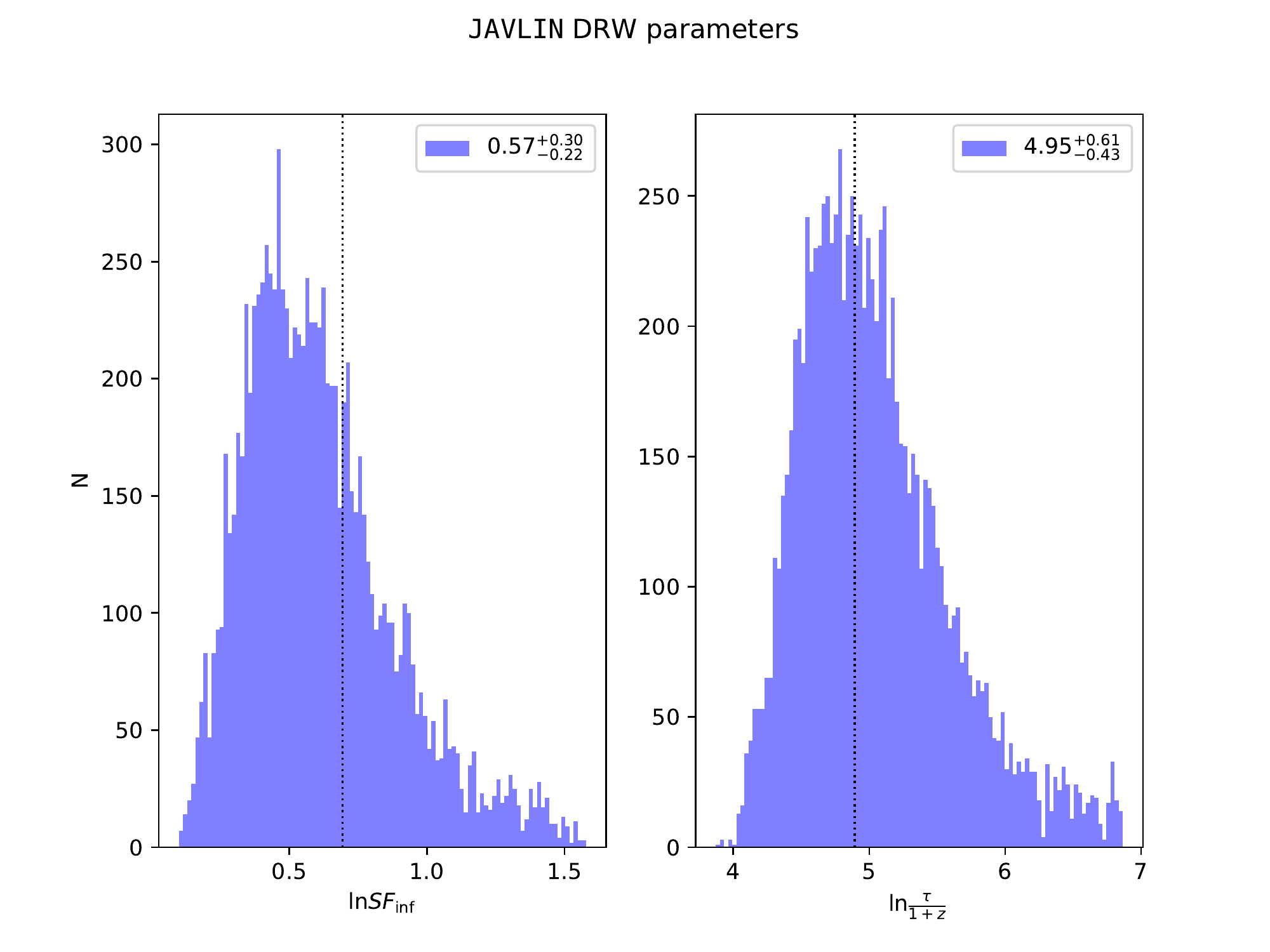}
\caption{
Posterior distributions of the DRW parameters from \JAVELIN: the structure function at infinity $SF_{\inf}$ ({\it left}) and the characteristic time-scale $\tau$ at the rest frame ({\it right}). 
The values on the top-right of each panel represent the 50th, 16th and 84th percentiles of the respective probability distribution.
The result shows that \JAVELIN\ can recover the input parameters $SF_{\inf}=2$ ($\ln SF_{\inf} =0.69$) and $\tau=200$ ($\ln \frac{\tau}{1+z}=4.89$), which are highlighted with dotted lines.
}
\label{fig:javelin_drw}
\end{figure*}

The posterior distribution of the DRW's parameters from \JAVELIN\ are shown in \fref{fig:javelin_drw}. 

In the case of \CREAM, we model the driving light curve with frequencies in the range 0.0005 to 0.4~cycles/day. The number of Fourier frequencies is $\sim800$. The time sampling both for light curves and for the transfer functions is 0.1~days. 

We set the black hole mass $\MBH=2\times10^8\Msun$ and the initial mass accretion rate is $\dot{M}=0.5\Msun/{\rm yr}$, varying with a step 0.01 $\Msun/{\rm yr}$.
We ignore the inner edge in \CREAM\ to match our simplified model. 
In practice, the position of the inner edge is outside the Schwarzschild radius, $R_s=2G\MBH/c^2$. 
A full account of the size of the inner edge is beyond the scope of this work.
The power law indices of the viscous and irradiation components of the temperature-radius profile are fixed to be $-3/4$.
We also fix the inclination to be 0 degree as the results do not change qualitatively by changing this parameter.

\begin{figure*}
\hspace*{-2cm}
\includegraphics[scale=0.72]{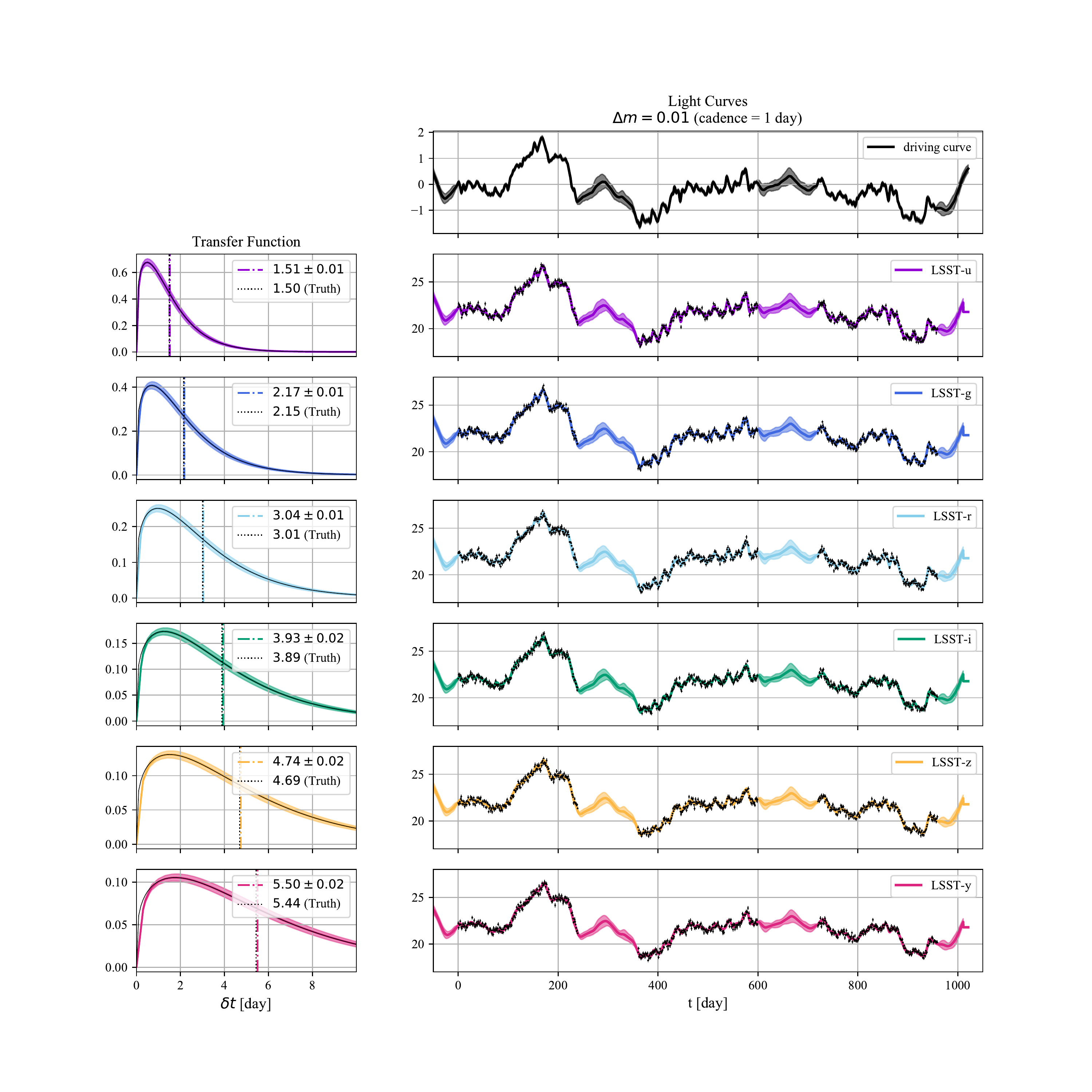}
\caption{{\it Left:} Transfer functions in each filter, as recovered with \CREAM, along with their means {\it in the observed frame} represented as dot-dashed vertical lines. 
The input transfer functions are shown as black curves and the means are labelled as black dot lines, which agrees well with the \CREAM's outputs.
{\it Right:} The inferred light curves from \CREAM, plotted as the mean and rms envelope of the MCMC samples, along with the simulated data, which are sampled with $\Delta m = 0.01$ mag over a period of 1000 days using a 1-day cadence and adding season gaps of 120 days every 240 days.
}
\label{fig:cream}
\end{figure*}

\begin{figure*}
\centering
\includegraphics[scale=0.8]{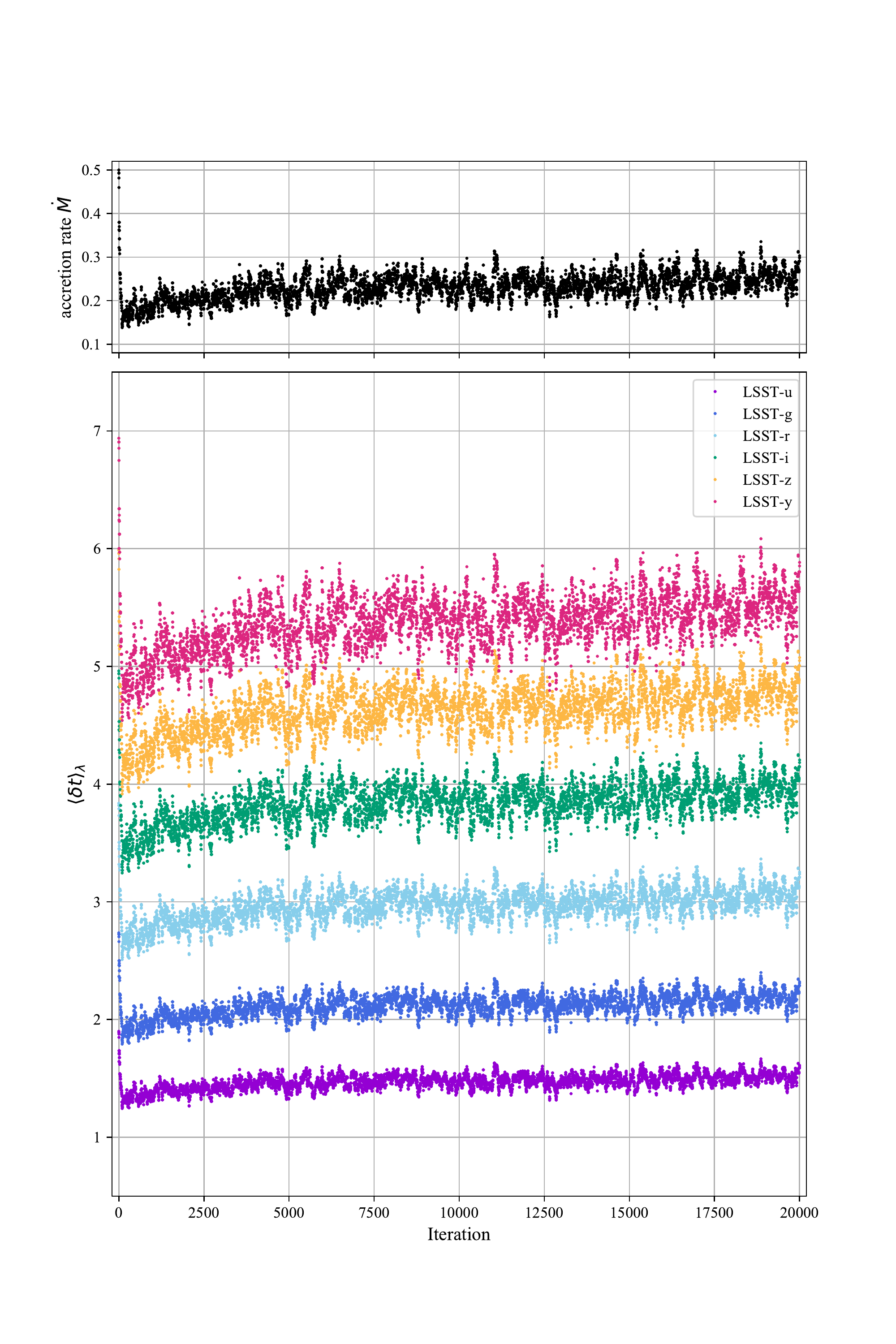}
\caption{
The samples for the accretion rate, $\dot{M}$, and the mean delays $\DTmean$ from \CREAM.
}
\label{fig:sample}
\end{figure*}

The MCMC fit of \CREAM\ is shown in \fref{fig:cream}.
The left column shows the inferred transfer functions with vertical lines showing the mean time delay.
The right column shows the response light curves in the LSST \ugrizy\ filters including $1\sigma$ uncertainty envelopes.
\CREAM\ outputs the posterior of accretion rate $\dot{M}$ and the mean delay of each transfer function $\DTmean$, and the samples. 
These are shown in \fref{fig:sample}.
We notice that in \fref{fig:cream} the light curve error envelopes do not sufficiently expand and contract in the seasonal gaps, indicating that the number of MCMC samples do not adequately cover the posterior parameter distribution, due to the computational limit on this high cadence and high accuracy data set.
As a consequence, the accuracy of the disk size measurement is insensitive to the number of MCMC samples, but some of the error distributions of the \CREAM\ parameters are likely to be underestimated and biased as shown in \fsref{fig:delay_measure} and \ref{fig:cdn_loss}.

One can convert $\dot{M}$ to the source size following the convention adopted in the \CREAM\ code:
\be
\begin{split}
R_{\lambda} & = 5.16\times10^{12} {\rm cm}\\
&\times \left[3+4\delta \eta (1-A)\right]^{1/3}
\left(\frac{\lambda}{\rm \mu m} \right)^{4/3}
\left(\frac{\MBH\dot{M}}{\Msun^2 {\rm yr}^{-1}} \right)^{1/3}
\end{split},
\ee
where $\delta=0$ is the ratio of lamp-post height and inner edge in Schwarzschild radius units $R_s=2GM/c^2$, and $A=0$ is the disk albedo. We note at this stage that users of \CREAM\ do not have access to all of the code parameters. 
We therefore use the mean output delays to infer the source size according to \eref{eqn:source_size}.
Although we use only $\LSSTu$ as a reference to estimate the disk size, we still find that using other filters as a reference leads to consistent result.
We fix the nominal error bars of the input light curves.



\end{document}